\documentclass[aps,prb,reprint,a4paper,floatfix,groupedaddress,amsmath,amssymb,amsfonts]{revtex4-1}
\usepackage{mathptmx}
\usepackage[scaled=0.9]{helvet}
\usepackage[utf8]{inputenc}
\usepackage{graphicx}
\usepackage{textcomp}
\usepackage{upgreek} 
\usepackage[dvipsnames]{xcolor}
\usepackage{soul} 
\usepackage{xspace}
\usepackage[rightcaption]{sidecap}

\usepackage[pdftex]{hyperref}
\hypersetup{
	pdftitle={Nature of excitons bound to inversion domain boundaries: Origin of the 3.45-eV band and its sub-structure in spontaneously formed GaN nanowires on Si(111)},
	colorlinks,
	citecolor=blue
	}

\usepackage[
,textwidth=17.5cm
,textheight=23.5cm
,verbose
,pdftex
]{geometry}

\newcommand{\degree}{$^{\circ}$\xspace}
\newcommand{\celsius}{$^{\circ}$C\xspace}
\newcommand{\dox}{$(D^{0},X_\text{A})$\xspace}
\newcommand{\doxA}{$(D^{0},X_\text{A})$\xspace}
\newcommand{\xA}{$X_\text{A}$\xspace}

\newcommand{\sfx}{$(I_{1},X)$\xspace}
\newcommand{\idb}{$(IDB^{*},X)$\xspace}
\newcommand{\idbl}{$(IDB^{*},X)_\text{L}$\xspace}
\newcommand{\idbh}{$(IDB^{*},X)_\text{H}$\xspace}
\newcommand{\IDB}{IDB$^{*}$\xspace}
\newcommand{\IDBs}{IDB$^{*}$s\xspace}
\newcommand{\Tgr}{$T_\text{S}$\xspace}

\setcitestyle{numbers,square}
\begin{document}
\makeatletter
\renewcommand\@biblabel[1]{\mbox{$\left[ \right.$}#1\mbox{$\left. \right]$}}
\makeatother


\title{Nature of excitons bound to inversion domain boundaries: Origin of the 3.45-eV luminescence lines in spontaneously formed GaN nanowires on Si(111)}

\author{Carsten Pfüller}
\email{pfueller@pdi-berlin.de}
\author{Pierre Corfdir}
\thanks{C.\,P. and P.\,C. contributed equally to this work.}
\author{Christian Hauswald}
\altaffiliation{Present address: DILAX Intelcom GmbH, Alt-Moabit 96b, 10559 Berlin, Germany.}
\author{Timur Flissikowski}
\author{Xiang Kong}
\altaffiliation{Previously at Paul-Drude-Institut für Festkörperelektronik.}
\author{Johannes K. Zettler}
\altaffiliation{Present address: LayTec AG, Seesener Str.~10--13,  10709 Berlin, Germany.}
\author{Sergio Fernández-Garrido}
\author{Pınar Doğan}
\altaffiliation{Present address: Department of Electrical and Electronics Engineering, Faculty of Engineering, Muğla Sıtkı Koçman University, Kötekli, 48000, Muğla, Turkey.}
\author{Holger T. Grahn}
\author{Achim Trampert}
\author{Lutz Geelhaar}
\author{Oliver Brandt}

\affiliation{Paul-Drude-Institut für Festkörperelektronik, Leibniz-Institut im Forschungsverbund Berlin e.V., Hausvogteiplatz 5--7, 10117 Berlin, Germany}


\begin{abstract}

We investigate the 3.45-eV luminescence band of spontaneously formed GaN nanowires on Si(111) by photoluminescence and cathodoluminescence spectroscopy. This band is found to be particularly prominent for samples synthesized at comparatively low temperatures. At the same time, these samples exhibit a peculiar morphology, namely, isolated long nanowires are interspersed within a dense matrix of short ones. Cathodoluminescence intensity maps reveal the 3.45-eV band to originate primarily from the long nanowires. Transmission electron microscopy shows that these long nanowires are either Ga polar and are joined by an inversion domain boundary with their short N-polar neighbors, or exhibit a Ga-polar core surrounded by a N-polar shell with a tubular inversion domain boundary at the core/shell interface. For samples grown at high temperatures, which exhibit a uniform nanowire morphology, the 3.45-eV band is also found to originate from particular nanowires in the ensemble and thus presumably from inversion domain boundaries stemming from the coexistence of N- and Ga-polar nanowires. For several of the investigated samples, the 3.45-eV band splits into a doublet. We demonstrate that the higher-energy component of this doublet arises from the recombination of two-dimensional excitons free to move in the plane of the inversion domain boundary. In contrast, the lower-energy component of the doublet originates from excitons localized in the plane of the inversion domain boundary. We propose that this in-plane localization is due to shallow donors in the vicinity of the inversion domain boundaries.

\end{abstract}

\maketitle

\section{Introduction}
\label{Sec1:introduction}

The promising optoelectronic properties of spontaneously formed GaN nanowires (NWs) reported in the pioneering works of \citet{Yoshizawa1997} and \citet{Sanchez-Garcia1998} have triggered world-wide research activities that have led to the demonstration of light-emitting \cite{Kikuchi2004} and light-harvesting devices \cite{Kamimura2013,Cansizoglu2015} based on group-III-nitride NWs. Despite this progress, several open questions still exist regarding the spontaneous formation of GaN NWs and their structural and optical properties. In particular, a prominent band at 3.45~eV has been widely reported in the low-temperature photoluminescence (PL) spectra of GaN NWs grown by plasma-assisted molecular beam epitaxy (PAMBE) on Si(111) \cite{Calleja2000,Robins2007b,Furtmayr_jap_2008,Corfdir_jap_2009,Brandt_prb_2010,Sam-Giao2013}. The origin of this band in GaN NWs has been a subject of a lively debate for almost two decades. In bulk GaN, two different recombination mechanisms are known to manifest themselves by luminescence lines at about 3.45~eV: first, the two-electron satellite (TES) of the donor-bound exciton transition [$(D^{0},X_\text{A})$] \cite{Skromme1996,Freitas2002,Wysmolek2002,Paskov2007} and, second, excitons bound to inversion domain boundaries \cite{Reshchikov2003}.

The intensity of the TES transitions in bulk GaN is about two orders of magnitude lower than that of the related \dox line \cite{Paskov2007}. In contrast, the 3.45-eV band in GaN NWs is often prominent and sometimes even dominates the near band-edge PL spectrum. Nevertheless, this band was ascribed to the TES by \citet{Corfdir_jap_2009}, who proposed that the distortion of the \dox wave function near the NW surface would lead to a strong enhancement of this transition. The same group substantiated this hypothesis by investigating the evolution of the 3.45-eV band with NW diameter and NW density \cite{Lefebvre2011}. However, investigations of the Fermi level pinning in GaN NWs \cite{Pfuller_prb_2010} as well as polarization-resolved PL and magneto-optical experiments \cite{Sam-Giao2013} later refuted the interpretation of the 3.45-eV band as an enhanced TES transition.

Inversion domain boundaries (IDBs) may give rise to intense PL lines in GaN films at 3.45~eV \cite{Schuck2001,Reshchikov2003,Kirste2011}. An IDB denotes a boundary between Ga- and N-polar GaN for which two different stacking sequences have been proposed by \citet{Northrup1996a}. The \IDB notation refers to the specific atomic structure at which each atom remains fourfold coordinated by exclusively forming Ga-N bonds across the boundary. In contrast, the unstarred IDB indicates a structure where the formation of Ga-Ga or N-N bonds would occur. The \IDB has an exceptionally low formation energy and does not induce electronic states in the band gap, thus facilitating the radiative recombination of excitons bound to these defects \cite{Reshchikov2003} (again in contrast to the unstarred IDB, which entails electronic states in the band gap). \citet{Robins2007b} suggested that the 3.45-eV band observed in PL spectra of GaN NWs is related to the presence of \IDBs in GaN NWs just as in the bulk. This suggestion, however, was not supported by investigations of the microstructure of GaN NWs at that time. In fact, transmission electron microscopy (TEM) performed on isolated GaN NWs invariably demonstrated the absence of extended defects in the NW volume \cite{Trampert2003,Bertness2006,Cerutti2006,Sekiguchi2007,Corfdir_jap_2009}.

Several groups therefore favored point defects as the origin for the 3.45-eV emission. In early work, in which GaN NWs were assumed to elongate axially via a Ga-induced vapor-liquid-solid mechanism, Ga interstitials were proposed as likely candidates for these point defects \cite{Calleja2000}. Later, GaN NW growth was understood to proceed under N-rich conditions, which were suggested to result in an enhanced formation of Ga vacancies near the NW surface \cite{Furtmayr_jap_2008}. \citet{Brandt_prb_2010} discussed the dependence of the 3.45-eV band on excitation density in terms of both planar defects such as the \IDB and abundant point defects and eventually favored the latter for being most consistent with the whole set of available data. However, the observation of strong NW-to-NW variations in the intensity of the 3.45-eV emission from single NWs \cite{Pfuller2010} contradicted this conclusion as well.

Recently, \citet{Auzelle2015b} presented definitive experimental evidence for the presence of \IDBs in GaN NWs grown on AlN-buffered Si(111). Subsequently, the same authors correlated \textmu-PL experiments with high-resolution scanning TEM performed on single GaN NWs grown on AlN-buffered Si \cite{Auzelle2015a}. They observed a systematic correlation between the presence of \IDBs in the GaN NW and transitions at 3.45~eV in its \textmu-PL spectrum and thus concluded that these transitions are caused by exciton recombination at \IDBs. Finally, they proposed that the 3.45-eV band is indicative for the presence of \IDBs in GaN NWs also for other substrates.

It is not uncommon to observe mixed polarities in AlN films grown on Si(111) \cite{Brubaker2011a}, and the coexistence of N- and Ga-polar NWs on such a film is therefore not an actual surprise. The spontaneous formation of GaN NWs directly on Si(111), however, is largely believed to occur on an amorphous SiN$_x$ interlayer formed during the (unintentional or intentionally promoted) nitridation of the Si substrate by the N plasma \cite{Trampert2003,Consonni2011}. The polarity of GaN NWs formed directly on such a nitridated Si(111) surface has been debated for a long time. In earlier work, the GaN NWs were mostly reported to be Ga polar \cite{Cherns2008,Furtmayr2008,Armitage2010,Lefebvre2011,Geelhaar_ieeejstqe_2011,Brubaker2011a}, while recent studies (which also include ensemble investigations with far better statitics) indicate that GaN NWs grow predominantly or even exclusively N polar \cite{Hestroffer2011,Schuster2012,Fernandez-Garrido2012,Carnevale2013a,Romanyuk2015,Minj2015a}. In any case, whether \IDBs form for GaN NWs on nitridated Si(111), and if so, at which density, are open questions.

In this paper, we report a comprehensive investigation of the structural and optical properties of GaN NWs on Si(111) fabricated with or without intentional substrate nitridation and within a wide range of substrate temperatures. This investigation focuses on the nature of the 3.45-eV band that is observed for all samples but with varying intensity. In Sec.~\ref{Sec2:experimental}, we briefly describe the samples under investigation and the experimental setups used for our study. Low-temperature cathodoluminescence (CL) spectroscopy and TEM are employed in Sec.~\ref{Sec3:CL} to investigate the origin of the 3.45-eV band. In accordance with the findings of \citet{Auzelle2015a}, we attribute this band to the  presence of \IDBs in our NWs. In Sec.~\ref{Sec4:PL}, we present an in-depth investigation of the 3.45-eV band comprising time-resolved and polarization-resolved PL spectroscopy accompanied by theoretical considerations based on data published by \citet{Fiorentini2003}. The observed doublet structure of the 3.45-eV band is identified to be due to the recombination of localized and delocalized states at the \IDB. The paper closes with a summary and conclusion in Sec.~\ref{Sec5:summary}.

\section{Experimental}
\label{Sec2:experimental}

The GaN NW ensembles studied in this work (see Tab.~\ref{Tab:samples} for an overview) were grown by PAMBE on AlN-buffered SiC(0001) (sample A) or Si(111) (samples B1--B7, C, and D). They were fabricated over the course of six years and in four different PAMBE systems (as indicated by the letters in their names). The Si(111) substrates were intentionally nitridated prior to NW growth for all samples except for sample D. We have intentionally chosen NW ensembles synthesized over a wide range of substrate temperatures \Tgr between 720 and 875~\celsius. All samples were grown under N-rich conditions with the Ga and N fluxes adjusted accordingly to account for the increased Ga desorption at elevated temperatures \cite{Fernandez-Garrido_jap_2009}. Due to the wide range of conditions, the different NW samples exhibited incubation times between a few minutes and hours as well as different NW densities, average NW diameters, and coalescence degrees. Further details on the growth conditions can be found elsewhere (sample A \cite{Fernandez-Garrido2012}, B1--B4 \cite{Dogan2011a}, B5--B7 \cite{Zettler2015}, C \cite{Consonni2011}, and D \cite{Cheze_apl_2010a}).

\begin{table}
\caption{List of the investigated samples. The substrate temperature \Tgr during growth, the year of fabrication, and the reference containing growth details are also given.}
\label{Tab:samples}
\begin{ruledtabular}
\begin{tabular}{ccccc}
sample & substrate & \Tgr (\celsius) & year & reference\\
A & AlN/SiC(0001) & 825\,\celsius & 2011 & \cite{Fernandez-Garrido2012}\\
B1 & nitridated Si(111) & 720\,\celsius & 2009 & \cite{Dogan2011a}\\
B2 & nitridated Si(111) & 780\,\celsius & 2009 & \cite{Dogan2011a}\\
B3 & nitridated Si(111) & 800\,\celsius & 2009 & \cite{Dogan2011a}\\
B4 & nitridated Si(111) & 820\,\celsius & 2009 & \cite{Dogan2011a}\\
B5 & nitridated Si(111) & 835\,\celsius & 2013 & \cite{Zettler2015}\\
B6 & nitridated Si(111) & 865\,\celsius & 2013 & \cite{Zettler2015}\\
B7 & nitridated Si(111) & 875\,\celsius & 2013 & \cite{Zettler2015}\\
C & nitridated Si(111) & 750\,\celsius & 2009 & \cite{Consonni2011}\\
D & Si(111) & 780\,\celsius & 2007 & \cite{Cheze_apl_2010a}\\
\end{tabular}
\end{ruledtabular}
\end{table}

The morphological and structural properties of the NWs were investigated by scanning electron microscopy (SEM) and TEM. For TEM, cross-sectional specimens were prepared by mechanical grinding, polishing, and subsequent Ar-ion polishing. The TEM images were recorded using an acceleration voltage of 300~kV. The polarity of the NWs was determined using convergent beam electron diffraction (CBED) complemented by subsequent dark-field TEM imaging.

SEM and CL spectroscopy were carried out in a field-emission instrument equipped with a CL system and a He-cooling stage. A photomultiplier tube was used for the acquisition of monochromatic images and a charge-coupled device (CCD) camera for recording CL spectra. Throughout the experiments, the acceleration voltage and the probe current of the electron beam were set to 5~kV and 0.75~nA, respectively. The spectral resolution amounted to 8~meV. Due to the strong quenching of the CL intensity of GaN NWs under the electron beam \cite{Campo_2004}, the irradiation time was kept to a minimum. Since the \dox transition is more prone to quenching than the 3.45-eV band \cite{Pfuller_prb_2010}, the bichromatic CL images shown in this work consist of two superimposed monochromatic false-color images recorded first at 3.47 and then at 3.45~eV. Control experiments performed in reverse order confirmed the validity of our results.

All PL experiments were performed in backscatter geometry with the samples mounted in liquid He-cooled cryostats offering continuous temperature control from 5 to 300~K. All measurements were conducted with NW ensembles except for the polarization-resolved PL experiments, where single NWs had been dispersed onto bare Si(111) wafers prior to the measurement. For continuous-wave PL spectroscopy, the 325-nm line of a HeCd laser was used to excite the samples. The laser was focused to a spot with a diameter of 1~\textmu m by a near-ultraviolet microscope objective with a numerical aperture of 0.65. The photoluminescence signal was collected by the same objective and dispersed by a spectrometer with an energy resolution of 0.25 to 1~meV. The dispersed signal was detected by a CCD camera. Polarization-resolved measurements were carried out using a half-wave plate followed by a linear polarizer \cite{Corfdir2015}. Time-resolved PL measurements were performed by focusing the second harmonic (325~nm) of an optical parametric oscillator pumped by a femtosecond Ti:sapphire laser (pulse width and repetition rate of 200 fs and 76~MHz, respectively). The PL signal was dispersed by a monochromator and detected by a streak camera operating in synchroscan mode. The energy and time resolutions are 2~meV and 20~ps, respectively.

\section{IDB$^*$\lowercase{s} as the origin of the 3.45-\lowercase{e}V band in G\lowercase{a}N NW\lowercase{s} on S\lowercase{i}(111)}
\label{Sec3:CL}

\begin{figure}
\includegraphics[width=\columnwidth]{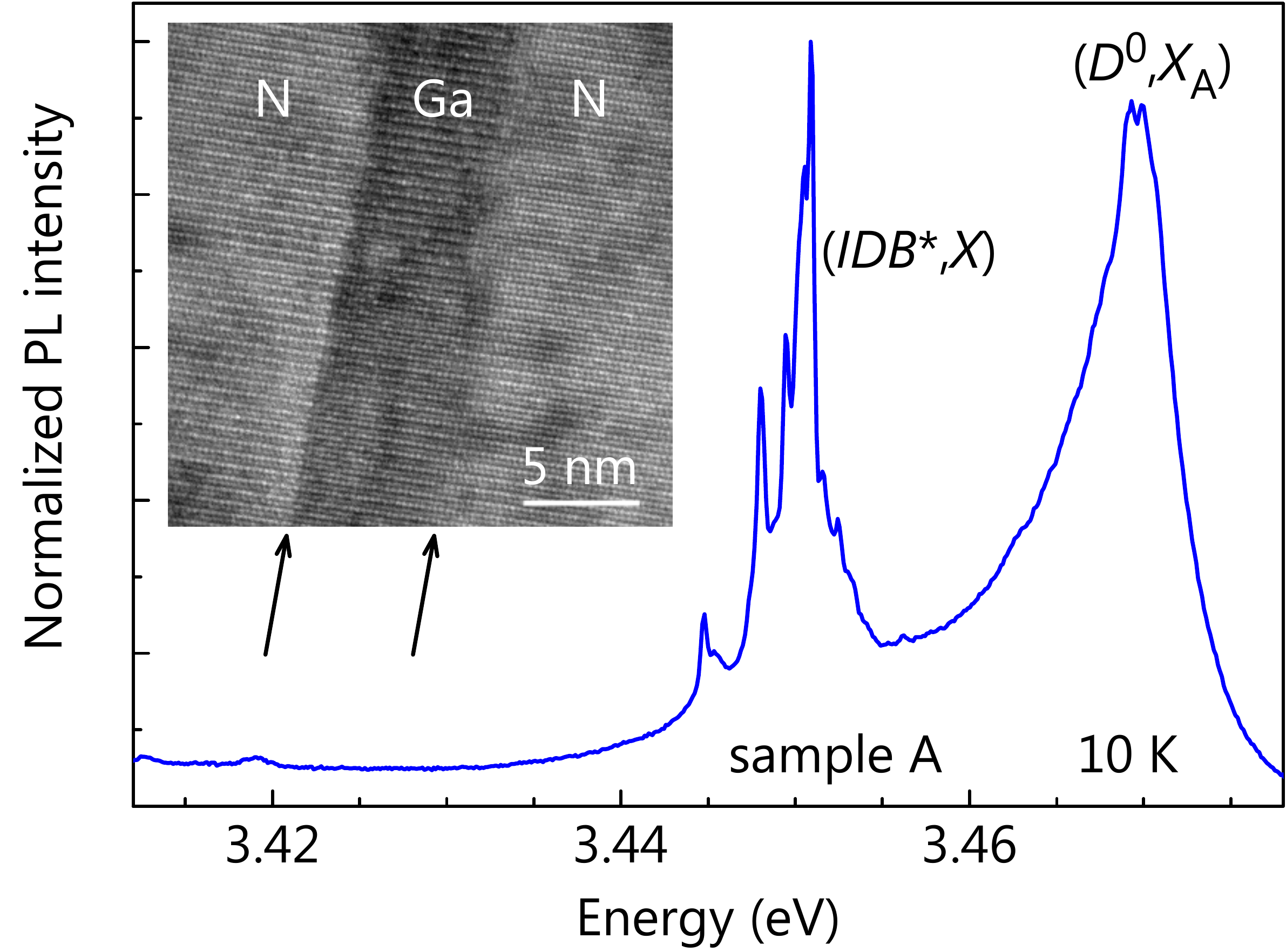}
\caption{Low-temperature (10~K) PL spectrum of a GaN NW ensemble grown on AlN/SiC(0001) (sample A). In addition to the \dox transition at 3.469~eV, several very narrow lines around 3.45~eV are observed. The high-resolution TEM image in the inset shows the central part of a GaN NW of sample A revealing a Ga-/N-polar core/shell structure. The arrows denote the \IDBs between the Ga-polar core and the N-polar shell.}
\label{Fig01-PL-TEM}
\end{figure}

In this section we will establish the correlation between \IDBs and the 3.45-eV band observed for GaN NWs on Si(111). Prior to a systematic investigation of our GaN NW ensembles on Si(111), we examine sample A which we already know to contain a very high density of IDBs or, as we will discuss later, most likely \IDBs. This sample was part of our study devoted to the role of substrate polarity in the formation of GaN NWs \cite{Fernandez-Garrido2012}. For this study, we attempted to induce the formation of GaN NWs on substrates with well-defined polarity, namely, AlN/SiC$(0001)$ and AlN/SiC$(000\bar{1})$. In the present work, we discuss further aspects of the sample synthesized at a substrate temperature of 825\,\celsius on SiC$(0001)$, which is here referred to as sample A. The polar nature of the SiC substrate is known to determine the polarity of group-III-nitride layers deposited on it \cite{Fernandez-Garrido2012}, and the AlN buffer was consequently found to be Al polar. Initiating the deposition of GaN at conditions typical for the growth of NWs resulted in a highly faceted Ga-polar GaN layer interspersed with sparse vertical NWs [SEM images of sample A can be found in Figs.~2(c) and 2(d) in Ref.~\onlinecite{Fernandez-Garrido2012}]. We have found the majority of these NWs to be N polar due to a Si-induced polarity flip at the interface between AlN and GaN. Consequently, \IDBs form upon coalescence between the Ga-polar matrix and the N-polar NWs \cite{Fernandez-Garrido2012}, and the sample is thus characterized by an exceptionally high fraction of NWs containing an \IDB.

\begin{figure}[b]
\includegraphics[width=\columnwidth]{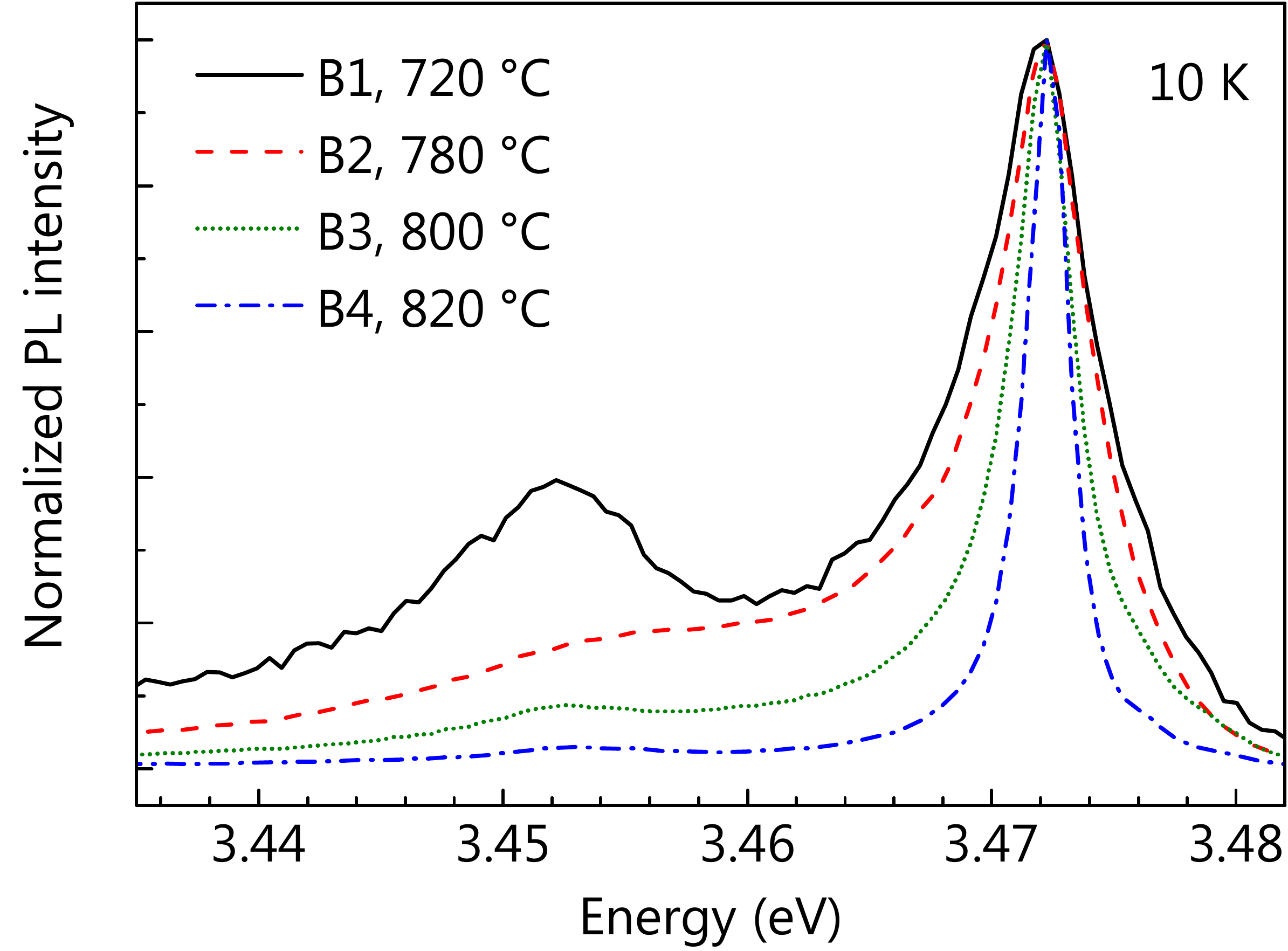}
\caption{Low-temperature (10~K) PL spectra of samples B1, B2, B3, and B4 grown on Si(111). The spectra are normalized to the \dox transition. The substrate temperature used for the growth of each sample is specified in the figure.}
\label{Fig02-T-series-norm}
\end{figure}

\begin{figure*}
\includegraphics[width=13cm]{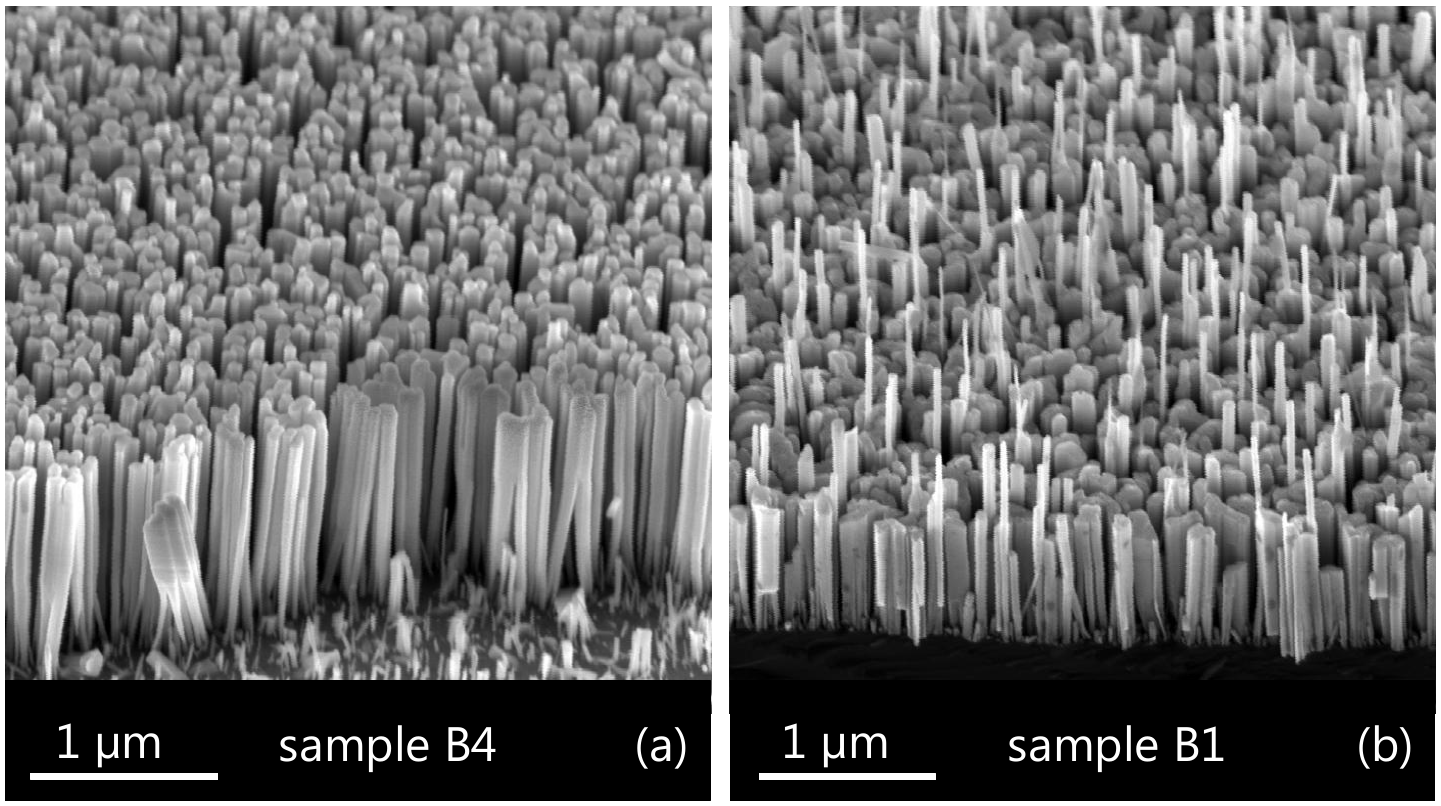}
\caption{Bird's eye view SEM images of samples (a) B4 and (b) B1 grown at $T_\text{S}=820$ and 720 \celsius, respectively. Samples grown at higher substrate temperatures such as the ones depicted in (a) exhibit uniform NW lengths. In contrast, for GaN NWs fabricated at lower substrate temperatures [cf. (b)], a dense matrix of short NWs is interspersed by a few long and thin ones.}
\label{Fig03-SEM}
\end{figure*}

The PL spectrum of sample A is shown in Fig.~\ref{Fig01-PL-TEM}. The \dox transition is observed at 3.469~eV, but the spectrum also exhibits several intense and narrow lines around 3.45~eV. The slight redshift of the \dox transition with respect to the one usually observed for GaN NWs (3.471~eV) as well as its comparatively large line width suggests that it mainly originates from the Ga-polar GaN layer. The strong transitions observed around 3.45~eV are consistent with the high fraction of NWs with an \IDB found in our previous analysis [see Fig.~4(b) in Ref.~\onlinecite{Fernandez-Garrido2012} for an example]. Moreover, further investigations of sample A by TEM conducted in the course of the present work revealed in addition the existence of Ga-/N-polar core/shell NWs analogous to those reported by \citet{Auzelle2015a} (see the high-resolution TEM image displayed in the inset of Fig.~\ref{Fig01-PL-TEM}). All these findings are in agreement with those reported in Ref.~\onlinecite{Auzelle2015a} and strongly suggest that the lines observed around 3.45~eV in Fig.~\ref{Fig01-PL-TEM} are in fact due to excitons bound to \IDBs. Note that we ascribe the corresponding transitions to the \IDB rather than to the unstarred IDB because of the former's low formation energy and particularly the absence of dangling bonds, promoting the radiative decay of the exciton bound to it. We will therefore label these transitions in all what follows as \idb. An interesting finding in this context is the fine structure of the 3.45-eV band visible in Fig.~\ref{Fig01-PL-TEM}. The origin of the distinct narrow lines in the PL spectrum will be addressed in Sec.~\ref{Sec4:PL}.

A prominent band at 3.45~eV is commonly also observed for GaN NWs grown on Si(111) \cite{Calleja2000,Robins2007b,Furtmayr_jap_2008,Corfdir_jap_2009,Brandt_prb_2010,Sam-Giao2013}. Figure~\ref{Fig02-T-series-norm} shows the evolution of the low-temperature (10~K) PL spectrum from GaN NW ensembles formed on Si(111) (samples B1--B4) under identical conditions except for the substrate temperature. The linewidth of the \dox transition decreases with \Tgr increasing from 720 to 820\,\celsius, indicating a progressive reduction of micro-strain induced by NW coalescence \cite{Jenichen2011a, Zettler2015, Fernandez-Garrido2014}. In parallel, the intensity of the 3.45-eV band with respect to that of the \dox line also decreases with increasing \Tgr. If this band is also related to the presence of \IDBs in GaN NWs grown directly on Si(111), its evolution seems to suggest that NWs fabricated at a higher \Tgr exhibit a lower density of \IDBs. In the following, we investigate whether this decrease in the relative intensity of the 3.45-eV band with increasing \Tgr is correlated with changes in the morphology of the NW ensemble. Furthermore, we examine the spatial distribution of the different spectral components for various NW ensembles.

\begin{figure*}
\includegraphics[width=13cm]{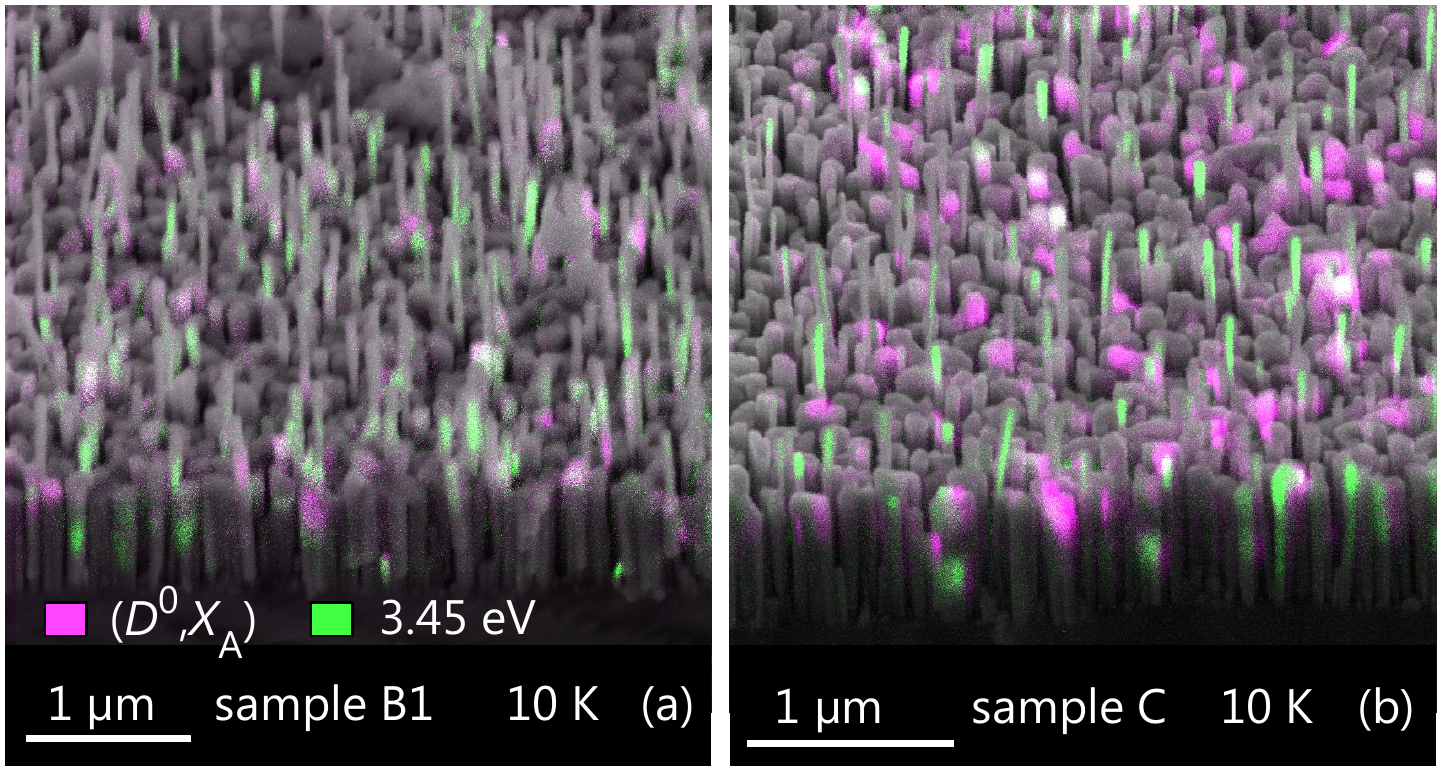}
\caption{Bird's eye view SEM images superimposed with bichromatic CL maps of samples (a) B1 and (b) C acquired at 10~K. NWs exhibiting an intense emission at 3.47 [corresponding to the \dox transition] and 3.45~eV appear magenta and green (linear intensity scale), respectively, while those emitting at both energies appear white.}
\label{Fig04-CL}
\end{figure*}

Figures~\ref{Fig03-SEM}(a) and \ref{Fig03-SEM}(b) display SEM bird's eye view images of samples B4 and B1, which were grown at $T_\text{S}=820$ and 720\,\celsius, respectively. The morphology of these samples is representative for NW ensembles grown at high and low \Tgr. For substrate temperatures higher than 750\,\celsius, the NW ensemble usually exhibits a NW density of about $5\times10^{9}$~cm$^{-2}$ together with a uniform height distribution as shown in Fig.~\ref{Fig03-SEM}(a) \cite{Sabelfeld2013}. In contrast, for temperatures significantly lower than this value, the ensemble morphology is characterized by a dense ($1\times10^{10}$~cm$^{-2}$) matrix of short and highly coalesced NWs interspersed by long NWs with a much lower density of $1\times10^{8}$~cm$^{-2}$ [cf.\ Fig.~\ref{Fig03-SEM}(b)]. Top-view SEM images (not shown here) indicate that most of the long NWs are attached to short NWs.

To identify a possible correlation between the peculiar morphology of sample B1 and its intense 3.45-eV band, we record the spatial intensity distribution of the two distinct PL bands by CL spectroscopy. Figure~\ref{Fig04-CL}(a) shows the superposition of an SEM image with a bichromatic CL map recorded at 3.47 and 3.45~eV from sample B1 at 10~K. Clearly, the 3.45-eV band (color coded in green) originates almost exclusively from long NWs, whereas the \dox line (color-coded in magenta) stems mostly from the top part of short NWs. Figure~\ref{Fig04-CL}(b) shows the same measurement for sample C, which is another typical representative for a GaN NW ensemble grown at comparatively low \Tgr. The spatial distribution of the emission at 3.47 and 3.45~eV is the same as for sample B1, as indeed for all samples exhibiting the morphology characteristic for growth at low substrate temperatures. Note that in Figs.~\ref{Fig04-CL}(a) and \ref{Fig04-CL}(b) many NWs appear not to emit at all (i.e., with the applied linear intensity scale they neither appear magenta nor green). This is consistently observed for samples grown at low \Tgr and implies that only a few NWs actually contribute to the CL signal. The strong quenching of the CL under electron irradiation furthermore enhances this phenomenon \cite{Lahnemann2016}.

\begin{figure}[b]
\includegraphics[width=\columnwidth]{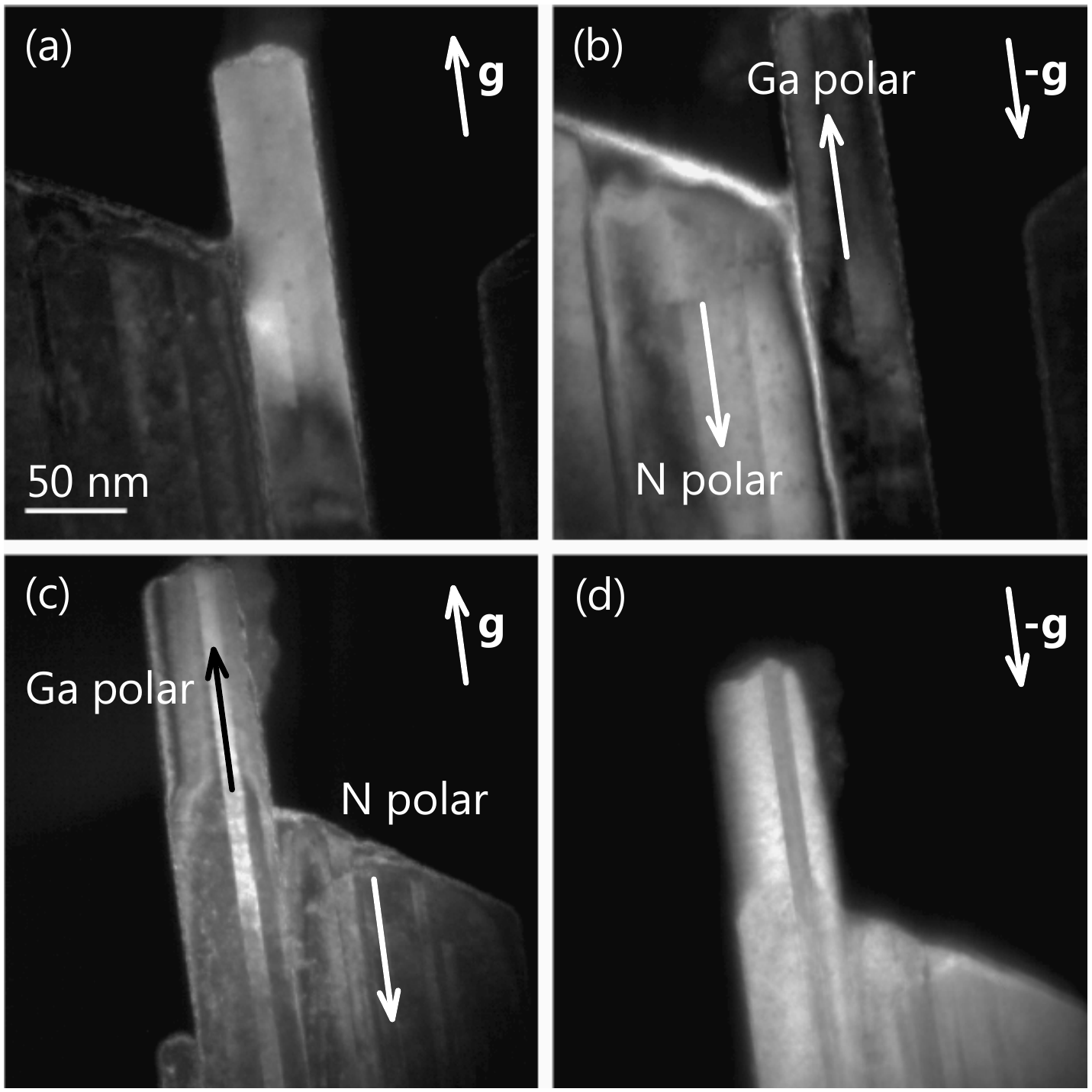}
\caption{Representative dark-field TEM images of sample C showing the columnar matrix of short N-polar NWs and the long NWs to be either [(a) and (b)] Ga polar or [(c) and (d)] to exhibit a Ga-/N-polar core/shell structure. An inverted diffraction vector $\mathbf{g}$ also inverts the contrast between the Ga- and N-polar material.}
\label{Fig05-TEM}
\end{figure}

\begin{figure*}
\includegraphics[width=\textwidth]{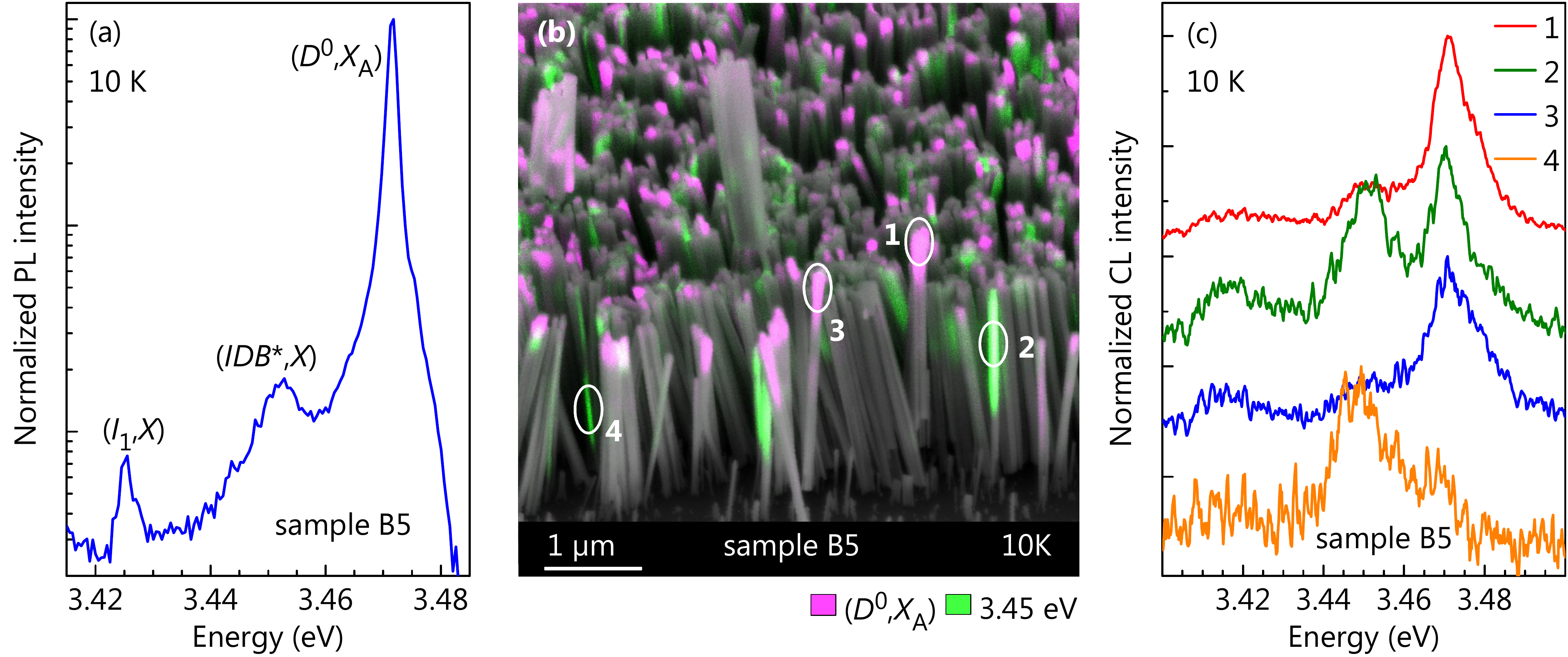}
\caption{(a) Low-temperature (10~K) PL spectrum of sample B5 grown at $T_\text{S} = 835$~\celsius. The line labeled \sfx is due to excitons bound to $I_1$ basal-plane stacking faults. (b) Bird's eye view SEM image of sample B5 superimposed with a bichromatic CL map acquired at 10~K. Magenta (green) areas imply that the related NWs emit dominantly at 3.471~eV (3.45~eV). The circles indicate the locations at which the spectra in panel (c) have been recorded. (c) Low-temperature (10~K) CL spectra of individual NWs of sample B5. The spectra have been recorded after acquisition of the CL map in panel (b) and have been shifted vertically for clarity.}
\label{Fig06_PL-CL}
\end{figure*}

To clarify the reason for this spatial distribution, we have performed CBED as well as TEM imaging on several NW samples grown at low \Tgr. Using CBED, it is straightforward to determine the polarity of the short NWs constituting the columnar matrix in these NW ensembles, since their effective diameter is large due to the high degree of coalescence. The polarity of this matrix was found to be exclusively N polar for both samples B1 and C (not shown here). For the long NWs with diameters below 50~nm, the polarity cannot be reliably determined by CBED. However, since we know the polarity of the columnar matrix, we can instead employ dark-field TEM and exploit the fact that opposite polarities induce a contrast inversion in images recorded by this technique. In addition, inverting the diffraction vector $\mathbf{g}$ should also invert the contrast for both polarities. Representative dark-field micrographs are displayed in Fig.~\ref{Fig05-TEM}. The opposite contrast between the long NW and its short neighbor in Fig.~\ref{Fig05-TEM}(a) recorded with $\mathbf{g}=0002$ as well as the contrast inversion in Fig.~\ref{Fig05-TEM}(b) recorded with $\mathbf{g}=000\bar{2}$ demonstrate that the long NW is Ga polar. The situation is more complex in Figs.~\ref{Fig05-TEM}(c) and \ref{Fig05-TEM}(d). Here, the shell of the long NW has the same polarity as the adjacent material, but it clearly has a Ga-polar core. These peculiar polarity core/shell NW structures seem to be identical to those observed in Fig.~\ref{Fig01-PL-TEM} and in Ref.~\onlinecite{Auzelle2015a}, suggesting that the mechanism giving rise to the formation of such structures is a general one and does not depend on the substrate.

The long NWs are either Ga polar [Figs.~\ref{Fig05-TEM}(a) and \ref{Fig05-TEM}(b)] or have a core/shell structure with a Ga-polar core surrounded by a N-polar shell [Figs.~\ref{Fig05-TEM}(c) and \ref{Fig05-TEM}(d)]. In the former case, planar \IDBs are formed at the coalescence boundaries between the long Ga-polar NWs and the N-polar columnar matrix. In the latter case, the Ga-/N-polar core/shell structure results in the formation of an \IDB tube.
\IDBs thus exist either at the junctions between the long NWs and the surrounding columnar matrix or directly within the long NWs themselves. It is thus very plausible that the 3.45-eV emission, which arises almost exclusively from the long NWs [see Fig.~\ref{Fig04-CL}], originates from the \idb complex.

\begin{figure}[b]
\includegraphics[width=\columnwidth]{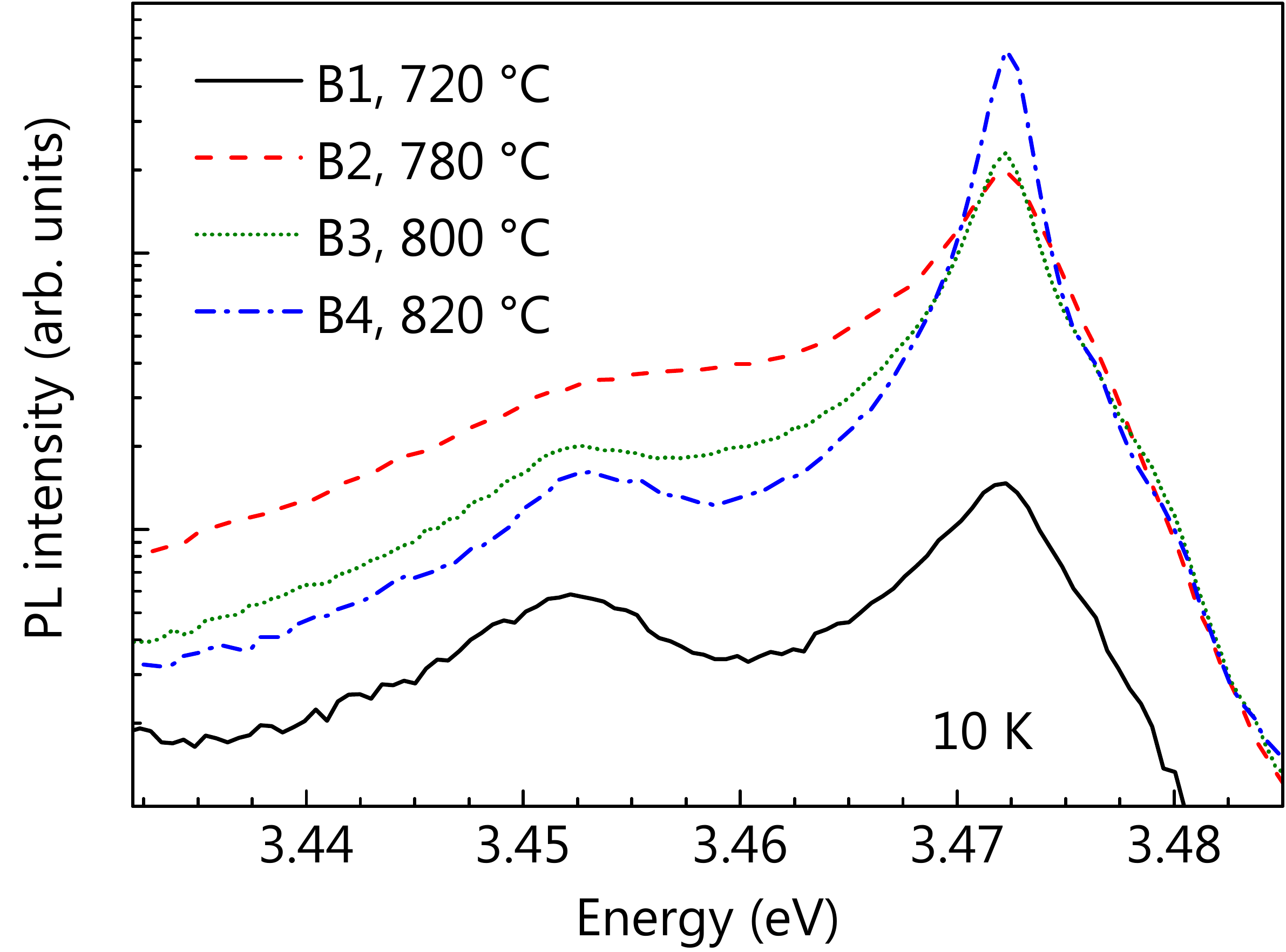}
\caption{Low-temperature (10~K) PL spectra of samples B1, B2, B3, and B4. The samples have been measured side-by-side under identical conditions, and the spectra are displayed on an absolute intensity scale. The substrate temperature used for the growth of each sample is specified in the figure.}
\label{Fig07-T-series-abs}
\end{figure}

We have shown so far that NW ensembles grown at a lower \Tgr exhibit both short and long NWs and that the optical transition at 3.45~eV is due to exciton recombination at \IDBs in long thin NWs. However, with increasing \Tgr, the NW ensembles are getting more homogeneous in diameter and length [Fig.~\ref{Fig03-SEM}(a)], and the intensity of the transition at 3.45~eV strongly decreases in comparison to that of the \dox as depicted in Fig.~\ref{Fig02-T-series-norm}. Figure~\ref{Fig06_PL-CL}(a) shows for sample B5 grown at $T_\text{S} = 835$\,\celsius that the intensity of the line at 3.45~eV for NW ensembles grown at high substrate temperatures can be two orders of magnitude smaller than that of the \dox line. In the bulk, the TES of the \dox is also centered at 3.45~eV and its intensity is about two orders of magnitude smaller than that of the \dox transition \cite{Paskov2007}. Consequently, the question arises whether the weak 3.45-eV band observed in Fig.~\ref{Fig06_PL-CL}(a) is related to \IDBs at all.

Figure~\ref{Fig06_PL-CL}(b) shows a bichromatic CL map of sample B5 and Fig.~\ref{Fig06_PL-CL}(c) depicts CL spectra taken at 10~K on individual NWs. Clearly, the \dox and the 3.45-eV emission lines do not coincide spatially, ruling out the standard two-electron satellites as a possible origin of the 3.45-eV band and suggesting instead that the 3.45-eV band also arises from the presence of \IDBs in these high-\Tgr NW ensembles. Remarkably, the density of NWs with dominant \idb transitions is comparable to that observed for low-\Tgr ensembles (cf.\ Fig.~\ref{Fig04-CL}), demonstrating that the density of \IDBs does not change significantly with substrate temperature. This finding seems to contradict the evolution of the relative intensity of the \idb band with \Tgr as depicted in Fig.~\ref{Fig02-T-series-norm}. However, plotting the same data on an absolute intensity scale as done in Fig.~\ref{Fig07-T-series-abs} reveals that the intensity of the \idb band is actually not significantly reduced with increasing \Tgr. The decrease of the relative intensity of the \idb band with increasing \Tgr is instead caused by the drastic increase in the \dox emission intensity, reflecting the reduced concentration of nonradiative point defects at high \Tgr \cite{Sobanska2015,Zettler2015}.

\section{Optical properties of the $\mathbf{(IDB^{*},X)}$ band}
\label{Sec4:PL}

We have established in the previous section that the 3.45-eV transition in GaN NWs on Si(111) is related to \IDBs regardless of the substrate temperature. In this section, we investigate the optical properties of excitons bound to \IDBs in more detail. In particular, we critically examine the consistence of our experimental results with the properties expected theoretically for this particular bound exciton state. We focus on samples which exhibit a detailed fine structure in their PL spectra.

\begin{figure}
\includegraphics[width=\columnwidth]{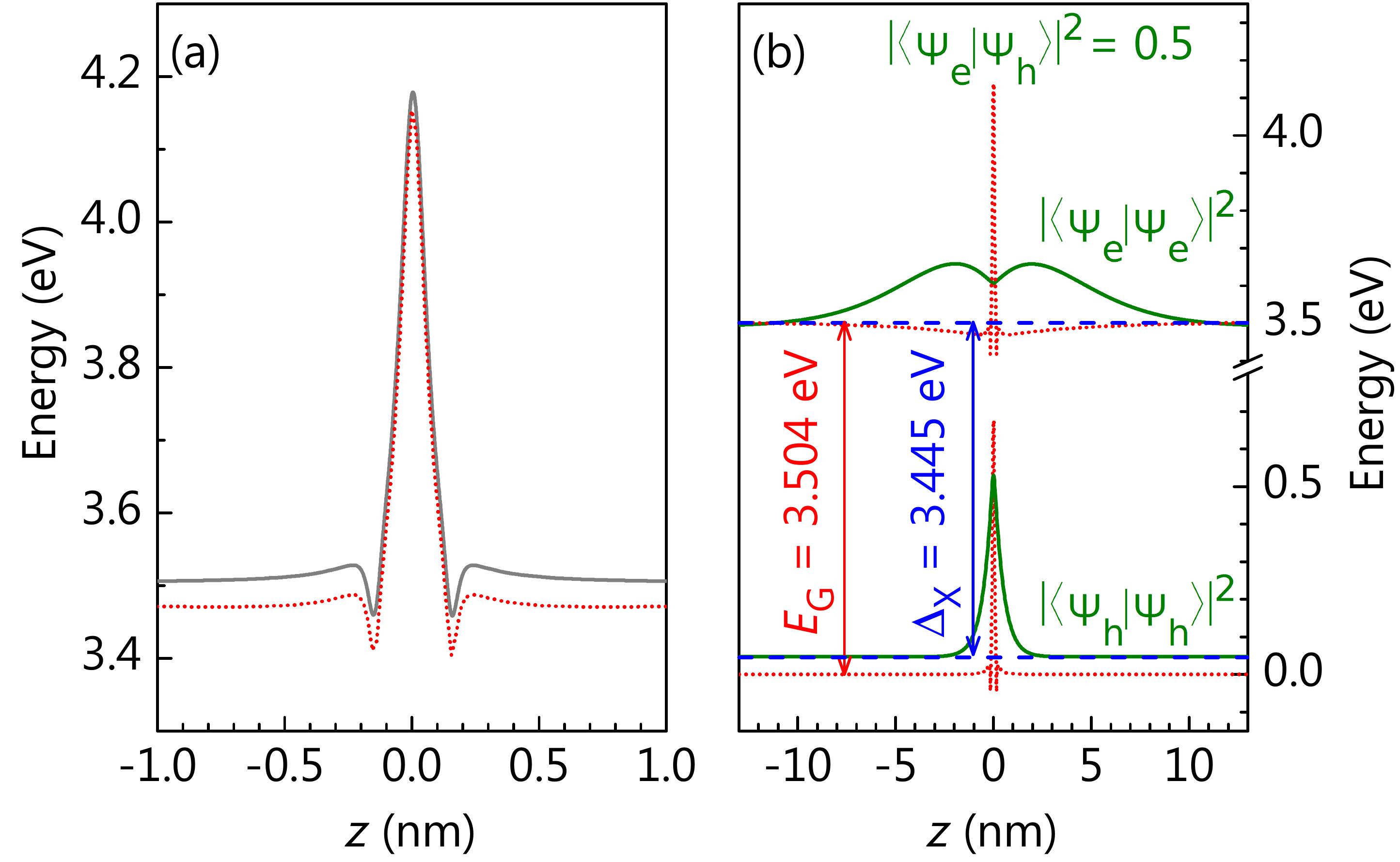}
\caption{(a) Electronic potential across an \IDB in GaN as calculated in Ref.~\onlinecite{Fiorentini2003} (solid line) and including the attractive potential exerted by the holes as derived from effective-mass calculations (dotted line). (b) Band profile (dotted lines), electron- and hole probability density (solid lines), and their energy levels (dashed lines) as a function of the distance to the \IDB as derived from effective-mass calculations [see (a)]. The values for the bandgap $E_\text{G}$, the transition energy $\Delta_\text{X}$, and the absolute square of the overlap integral between the electron and hole wavefunctions $|\langle\Psi_\text{e}|\Psi_\text{h}\rangle|^2$ are also given.}
\label{Fig08_Eff-Mass}
\end{figure}

One basic property of the \idb transition can be deduced directly from the fact that the \IDB is a planar defect, which laterally extends over several tens of nm and vertically spans the entire NW length. In analogy to the result in Ref.~\onlinecite{Corfdir2016} for basal-plane stacking faults, we hence expect the density of states of the \idb to be two-dimensional. As a result of the large number of states available, the \idb transition should be difficult to saturate even for high excitation conditions. Indeed, the \idb transition has been observed to scale linearly with excitation density even after the higher-energy \dox transition has started to saturate \cite{Reshchikov2003,Calleja2000,Brandt_prb_2010}.

Using density-functional theory, \citet{Fiorentini2003} calculated the electronic potential in the vicinity of an \IDB based on the stacking sequence proposed by \citet{Northrup1996a}. This potential, depicted in Fig.~\ref{Fig08_Eff-Mass}, acts as a barrier for electrons and as a quantum well for holes: an \IDB can therefore be seen as a type-II quantum well that binds holes. Electrons are maintained in the surrounding of the \IDB due to the Coulomb interaction with the holes. In addition, since the type-II quantum well formed by the \IDB is extremely thin, the electron and hole wavefunctions have significant amplitudes within and outside of the quantum well, respectively, leading to a large electron-hole overlap and thus making the radiative recombination of the \idb an efficient process. To get quantitative information on the properties of the electronic state associated with the \IDB, we calculated the wavefunction and the energy of an exciton in the potential profile obtained by \citet{Fiorentini2003} using the variational approach described in Ref.~\onlinecite{Corfdir2012}. The result of our calculations is shown in Fig.~\ref{Fig08_Eff-Mass}. The calculations yield an electron-hole overlap, defined as the absolute square of the overlap integral between the electron ($\langle \Psi_e|$) and hole ($| \Psi_h \rangle$) wavefunctions, of $|\langle\Psi_\text{e}|\Psi_\text{h}\rangle|^2 = 0.5$. Therefore, despite the type-II band alignment across the \IDB plane, the \idb transition possesses a large oscillator strength in agreement with the high intensity observed experimentally [cf.\ Figs.~\ref{Fig04-CL} and \ref{Fig06_PL-CL}(b)]. For isotropic electron and hole masses of $0.2 m_0$  and $1.0 m_0$,\cite{Vurgaftman2003} with $m_0$ denoting the mass of the free electron, the energy of this transition $\Delta_\text{X}$ is found to be 3.445~eV, i.\,e., close to the experimental value.

\begin{figure}
\includegraphics[width=\columnwidth]{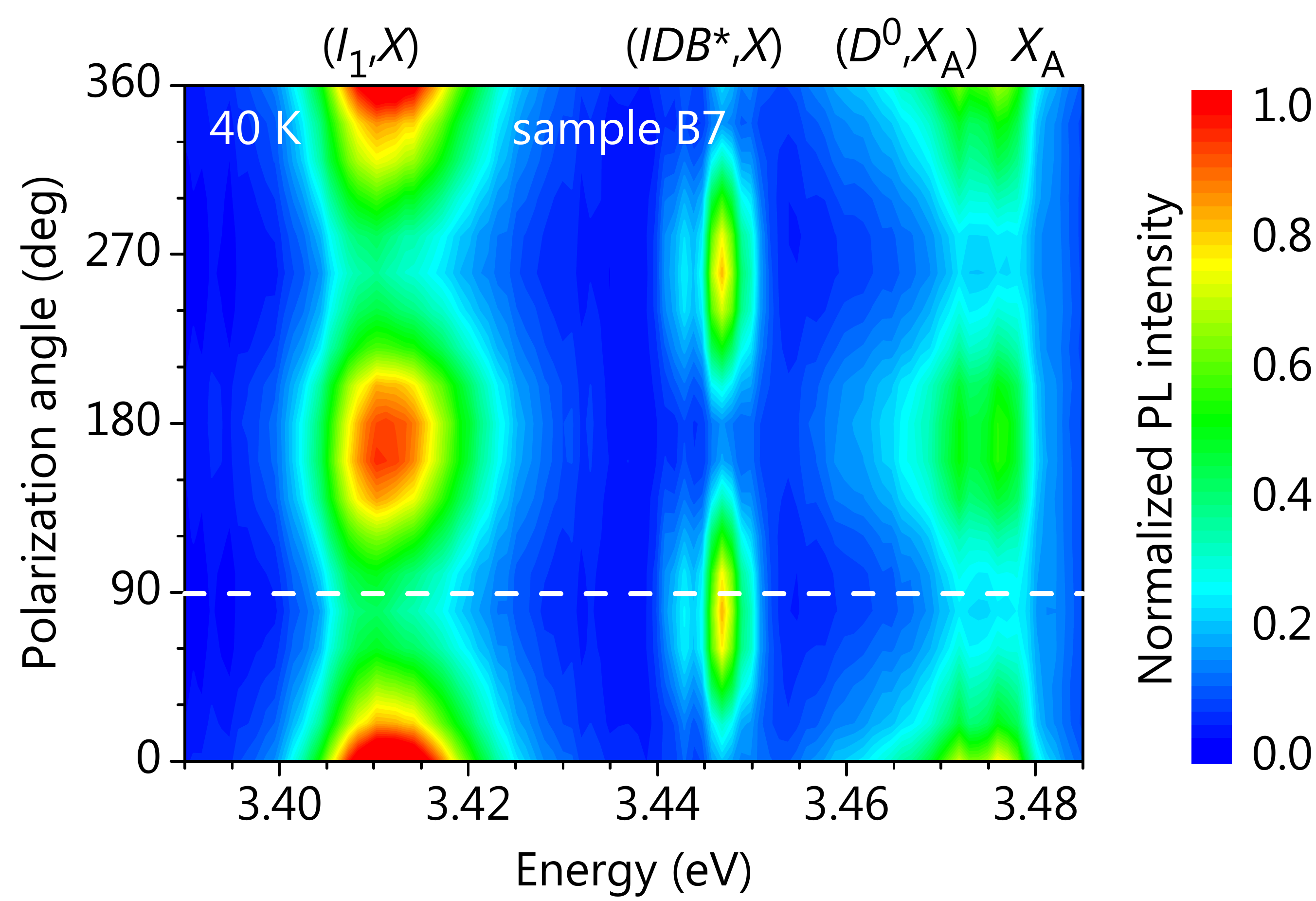}
\caption{Polarization map of the near-band-edge luminescence of a dispersed NW of sample B7 recorded at 40~K. The NW axis, which is parallel to the $c$ axis of GaN, is oriented along 90\degree as indicated by the dashed line.}
\label{Fig09_Polarization}
\end{figure}

\citet{Fiorentini2003} also predicted a significant mixing between the $\Gamma_{7+}^v$ and $\Gamma_9^v$ valence bands due to the \IDB. To verify this prediction, we have performed polarization-resolved PL experiments at 40~K on single NWs from sample B7 dispersed on a Si substrate. A typical polarization-resolved PL map taken on a single NW is shown in Fig.~\ref{Fig09_Polarization}. In agreement with previous reports \cite{Sam-Giao2013,Corfdir2015}, the free A exciton (\xA) and the \doxA transitions are polarized perpendicular to the NW axis ($\perp c$), whereas the \idb band is polarized parallel to the NW axis ($\parallel c$). The three lowest optical transitions in strain-free bulk GaN obey selection rules such that the $\Gamma_{7}^c\times\Gamma_{9}^v$ transition is allowed only for light polarized $\perp c$, the $\Gamma_{7}^c\times\Gamma_{7+}^v$ transition for both light polarized $\parallel c$ and $\perp c$ , and the $\Gamma_{7}^c\times\Gamma_{7-}^v$ transition mostly for light polarized $\parallel c$. The \idb band is polarized $\parallel c$ as evident in Fig.~\ref{Fig09_Polarization}, thus suggesting that it originates from a pure $\Gamma_{7}^c\times\Gamma_{7-}^v$ transition, i.\,e., the C exciton \cite{Misra2007a}. For a NW with sub-wavelength diameter, however, we have to bear in mind that the dielectric contrast strongly suppresses emission polarized perpendicular to the NW axis and thereby artificially enhances the component polarized $\parallel c$ \cite{Corfdir2015}. The strong polarization of the \idb band along the NW axis ($\parallel c$) thus suggests a reversal in the order of the $\Gamma_{7}^v$ and $\Gamma_9^v$ valence bands in the \IDB, in agreement with the theoretical result of Fiorentini \citep{Fiorentini2003}. Note also that the \sfx band is polarized $\perp c$, demonstrating that the opposite behavior observed for the \idb transition is a consequence of the peculiar potential induced by the \IDB and not a characteristic of excitons bound to planar defects in general.

\begin{figure}
\includegraphics[width=\columnwidth]{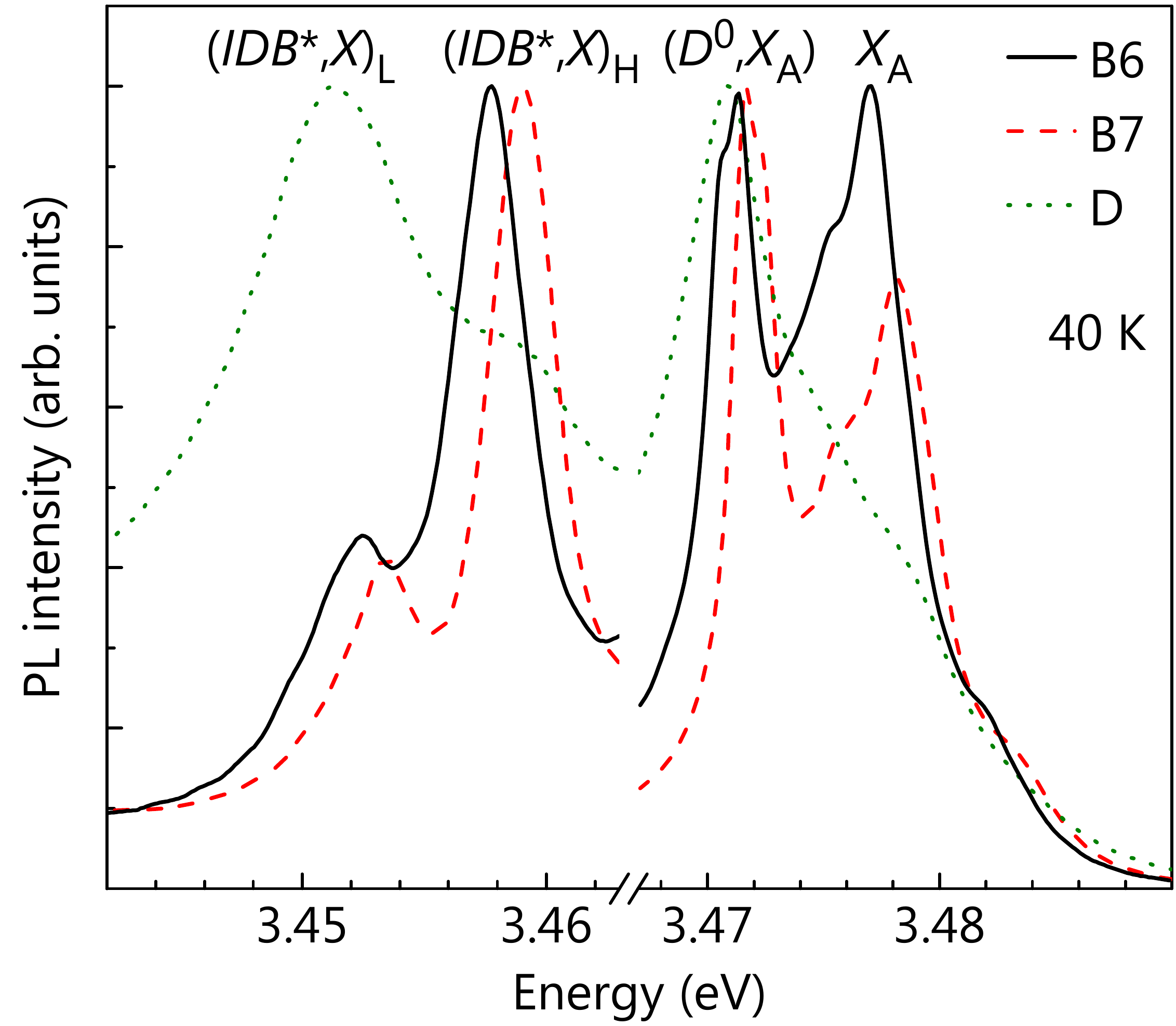}
\caption{(a) Low-temperature (40~K) PL spectra of sample B6, B7, and D grown at 865, 875, and 780\,\celsius, respectively. The spectral windows containing the \idb and the \dox have been normalized independently to the strongest transition.}
\label{Fig10-doublets}
\end{figure}

\begin{figure*}
\includegraphics[width=13cm]{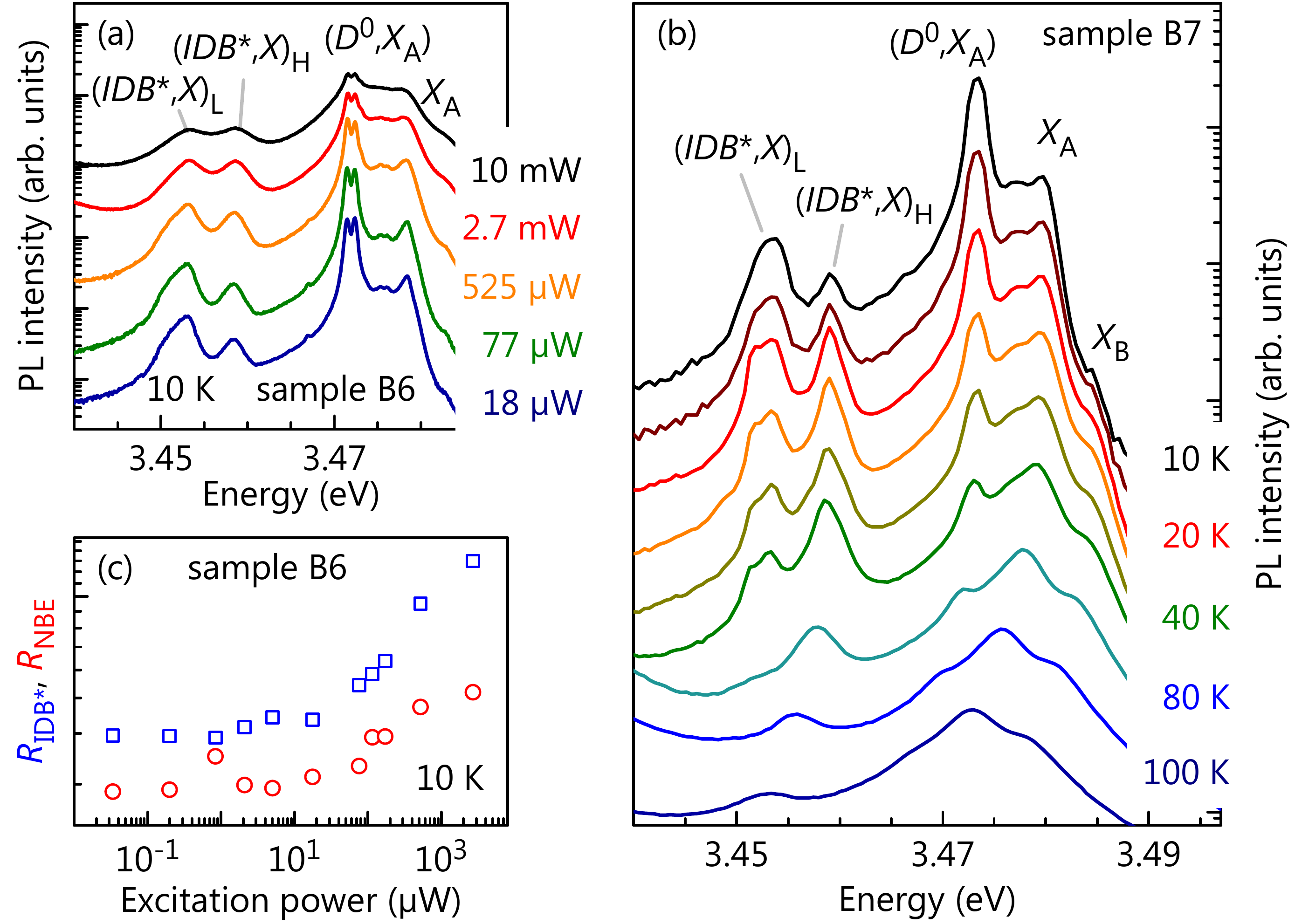}
\caption{(a) Low-temperature (10~K) PL spectra  of sample B6 for various excitation powers. (b) Evolution of the PL spectra of sample B7 with temperature. The spectra have been shifted vertically for clarity. (c) Intensity ratio $R_\text{\IDB}$ between the \idbh and and \idbl lines (squares) and $R_\text{NBE}$ between the \xA and \doxA transitions (circles) as a function of excitation power. The intensities have been obtained from a fit of the spectra in (a) with Voigt functions.}
\label{Fig11_T-P-Dependencies}
\end{figure*}

The reordering between $\Gamma_7^v$ and $\Gamma_9^v$ states proposed in Ref.~\onlinecite{Fiorentini2003} and observed in Fig.~\ref{Fig09_Polarization} is also consistent with the magneto-optical behavior of the \idb transition reported in Ref.~\onlinecite{Sam-Giao2013}. The Land\'{e} factor $g$ was found to be close to zero for the \idb, while for the \doxA transition in GaN NWs it was observed to be 1.75, a value similar to that reported for the bulk \cite{Sam-Giao2013}. As both the electron and hole Land\'{e} factors are extremely sensitive to valence band mixing \cite{Roth1959}, it is in fact not surprising to measure very different values of $g$ for the \doxA and the \idb lines.

All properties of the \idb transition discussed so far are consistently described by the electronic potential computed by \citet{Fiorentini2003}. However, this model does not provide any explanation for the fact that the \idb band often exhibits a doublet structure. As shown first by \citet{Calleja2000}, the band at 3.45~eV actually consists of two lines centered at about 3.449 and 3.455~eV \cite{Lefebvre2014,Corfdir2015}. This finding is confirmed by the PL spectra of samples B6, B7, and D at 40~K as shown in Fig.~\ref{Fig10-doublets}. For these three samples, the lower and higher energy component of the doublet is centered at about 3.452 and 3.458~eV, respectively. Note that the splitting between these two lines is almost identical to that between the \dox and the \xA transitions.

Figure~\ref{Fig11_T-P-Dependencies} shows the evolution of the \idb doublet with increasing excitation density and temperature for sample B6 and B7, respectively. At low temperatures and excitation densities, the \idb doublet is dominated by the lower energy transition [$(IDB^{*},X)_\text{L}$] at 3.452~eV. However, with increasing excitation density [Fig.~\ref{Fig11_T-P-Dependencies}(a)] or temperature [Fig.~\ref{Fig11_T-P-Dependencies}(b)], the higher energy transition [$(IDB^{*},X)_\text{H}$] takes over and eventually dominates the PL spectrum. At around 40~K, carriers start to escape from the \IDB, which manifests itself in a quenching of the \idb doublet \cite{Corfdir2014b}. As noted in Ref.~\onlinecite{Corfdir2015}, the excitation power and temperature dependences of the \idb doublet are similar to the ones observed for the \doxA and \xA transitions. In view of the fact that \IDBs act as quantum wells, we thus attribute the high-energy line of the \idb doublet to the recombination of excitons free to move along the \IDB plane and the low-energy one to excitons localized within this plane.

Localization within the plane of a quantum well usually occurs at well width fluctuations or due to alloy disorder \cite{Weisbuch1981}, which clearly cannot be the origin of the intra-\IDB localization observed here. Following the results reported for the localization of excitons in I$_1$ basal-plane stacking faults \citep{Corfdir2009c}, we propose that the short-range potential of shallow donors such as Si and O distributed in the vicinity of the \IDB induce the localization of excitons within the \IDB plane. The excitation dependence of the intensity ratio at 10~K between the \idbh and \idbl lines with that between the \xA and the \doxA in Fig.~\ref{Fig11_T-P-Dependencies}(c) is consistent with this idea. Both ratios remain nearly constant for low excitation powers, but increase together for powers higher than 18~\textmu W. This finding indicates that the density of localized states within the \IDBs is comparable to the equivalent density of donors in NWs and thus confirms that intra-\IDB exciton localization occurs due to donors. Note that we computed the characteristic extent of the exciton wavefunction perpendicular to the \IDB plane to be 5~nm. Assuming that the diffusion length of excitons in an \IDB is 100~nm \cite{Corfdir2011,Nogues2014}, a donor density of 10$^{16}$~cm$^{-3}$ is found to be sufficient to localize excitons within the \IDB plane, a value close to those reported for unintentionally $n$-doped GaN NWs \cite{Pfuller2010}.

In ensemble measurements, the \idbl band typically has a line width of several meV, whereas single NWs exhibit numerous sharp lines in this region \cite{Brandt_prb_2010,Corfdir2015}. The small differences in transition energies originate from the varying distances between the involved donor and the \IDB plane \cite{Corfdir2009c}. This effect can also be seen very clearly for sample A in Fig.~\ref{Fig01-PL-TEM}. Due to the low NW density of $5\times10^8$~cm$^{-2}$ of this sample, only a small number of NWs is probed simultaneously, and the individual narrow lines can be resolved even in an ensemble measurement. The line width of the sharp \idbl lines in Fig.~\ref{Fig01-PL-TEM} and also in Fig.~3 of Ref.~\onlinecite{Corfdir2015} is resolution limited, demonstrating that these lines stem from the radiative decay of bound excitons. In contrast, ensemble spectra such as the ones shown in Figs.~\ref{Fig10-doublets} and \ref{Fig11_T-P-Dependencies} contain contributions from about $10^3$ NWs, and the individual transitions can no longer be resolved, but blend together to an \idbl band with a line width of several meV \cite{Brandt_prb_2010}.  Note that \citet{Schuck2001} observed a fine structure of the \idb band in bulk GaN already in 2001. The \idbh band exhibits a linewidth of a few meV as expected for delocalized states.

\begin{figure}[b]
\includegraphics[width=\columnwidth]{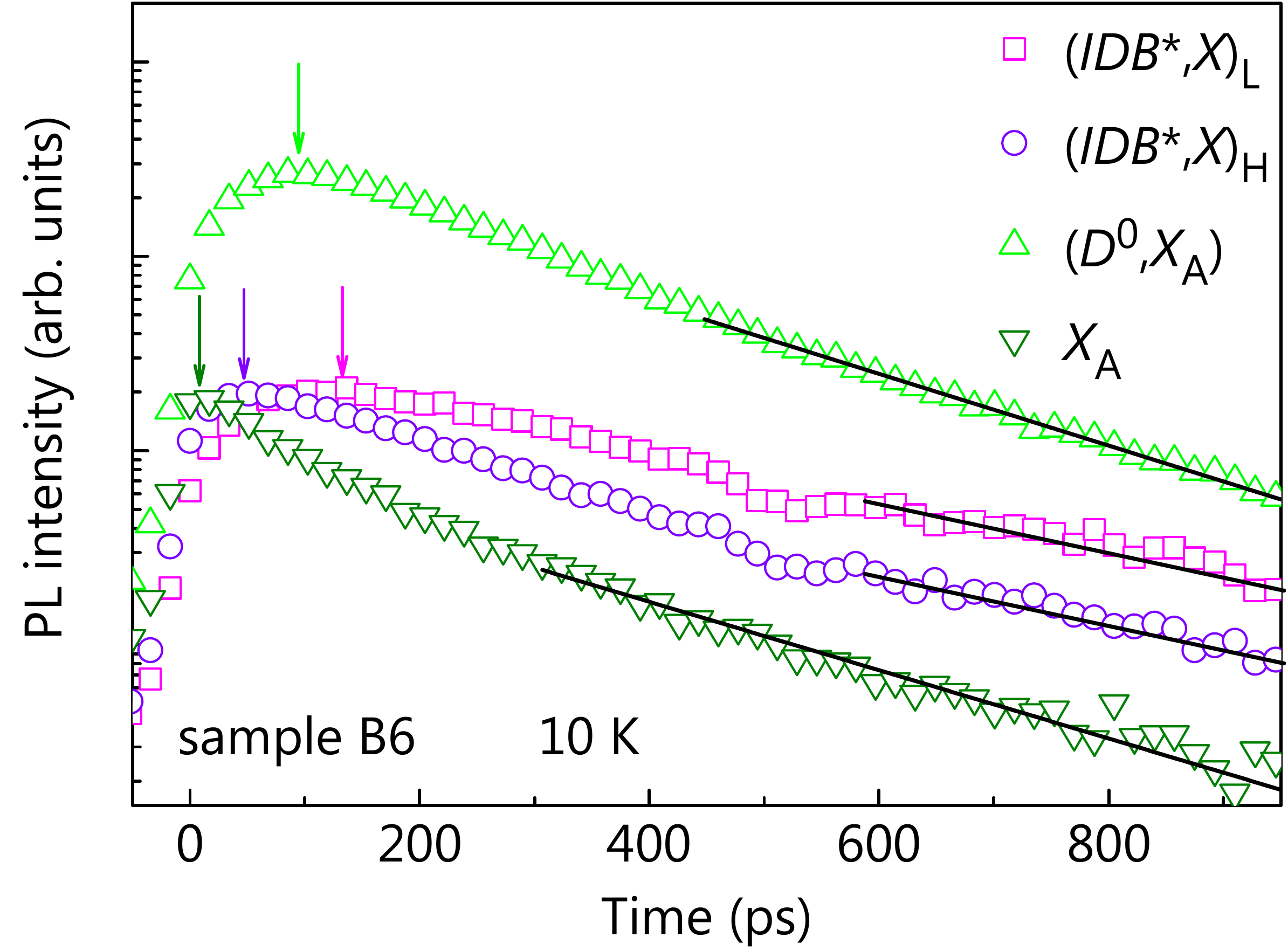}
\caption{Low-temperature (10~K) PL transients of the \doxA and \xA transitions as well as of the \idbl and \idbh lines of sample B6. The arrows denote the respective rise times, and the solid lines are exponential fits.}
\label{Fig12_TRPL}
\end{figure}

Finally, we have performed time-resolved PL experiments to obtain quantitative information on the capture efficiency of excitons by \IDBs and on the intra-\IDB localization process. Figure~\ref{Fig12_TRPL} shows PL transients of sample B6 measured at 10~K for the \xA and \doxA transitions as well as for the \idbh and \idbl lines. While the initial increase of the \xA PL intensity is almost instantaneous, it takes 90~ps for the \doxA PL intensity to reach a maximum. The latter time corresponds roughly to the characteristic time for the capture of excitons by donors with a density of about $5\times 10^{16}$~cm$^{-3}$ \cite{Hauswald2013}. In agreement with the results reported in Ref.~\onlinecite{Hauswald2013}, the \xA and \doxA PL decay in parallel at longer times as a result of the quasi-thermalization between these exciton states. Using an exponential fit, the effective decay time for the \xA and \doxA is found to be 195~ps, very similar to values obtained in previous reports \cite{Corfdir2009,Hauswald2014a,Sobanska2015}. This decay is much faster than expected for the radiative decay of the \doxA complex and is due to the nonradiative decay of the \xA at point defects \cite{Hauswald2014a}.

Analogously to the \dox and \xA transitions, the initial increase of the \idbl intensity is delayed compared to that of the \idbh. Interestingly, the rise time of the \idbh intensity is only 47~ps. Therefore, the capture of excitons by \IDBs is more efficient than their trapping by neutral donors due to the \IDBs' large capture cross-section. This situation differs from observations for the rise time of the \sfx transition at 10~K \cite{Corfdir2014b}, which is limited by the inefficient transport of excitons from one donor to the next \cite{Corfdir2009}. For time delays longer than 600~ps, the \idbl and \idbh decay in parallel, again demonstrating quasi-thermalization between these states. Finally, for time delays longer than 1~ns, the PL decay of the coupled \xA and \doxA states slows down and asymptotically approaches that of the \idb doublet (not shown), indicating full thermalization between the \xA, \doxA and \idb states. This process is the origin of the biexponential PL decay reported for the \doxA in Refs.~\onlinecite{Corfdir_jap_2009,Hauswald2013}.

\section{Summary and Conclusions}
\label{Sec5:summary}

The 3.45-eV band observed in low-temperature PL and CL spectra of spontaneously formed GaN NWs on Si(111) arises from planar or tubular \IDBs. While the former are due to the coalescence of adjacent Ga- and N-polar NWs, the latter form at the interface of Ga-/N-polar core/shell NWs. The intensity ratio between the \idb and the \dox transitions decreases with increasing \Tgr. This decrease is a consequence of the reduction in the density of nonradiative point defects with increasing \Tgr, leading to an increase in the absolute intensity of the \dox line. In contrast, the absolute intensity of the \idb band is neither directly governed by \Tgr nor by the presence or absence of an intentional nitridation step. The same applies to the abundance of \IDBs observed in spatially resolved, bichromatic CL maps.

These results confirm the idea of \citet{Auzelle2015a} that the 3.45-eV luminescence band in GaN NWs signifies the presence of \IDBs regardless of the substrate. In fact, we now know with certainty that \IDBs occur in GaN NW ensembles synthesized by PAMBE on AlN-buffered Si(111) \cite{Robins2007b,Auzelle2015a}, on nitridated Si(111), and on SiC(0001). The exceptionally low formation energy of \IDBs is obviously an important factor promoting their frequent occurrence. However, a prerequisite for the formation of an \IDB is the simultaneous presence of both Ga- and N-polar material. At present, it remains entirely unclear why an apparently constant fraction of the GaN nuclei on all of these different substrates are Ga polar. We also do not understand how Ga-polar NWs can evolve from these nuclei despite our inability to synthesize them intentionally on cation-polar substrates \cite{Fernandez-Garrido2012} and at substrate temperatures at which GaN(0001) usually decomposes. Finally, it is unclear how the peculiar Ga-/N-polar core/shell NWs form, which have been observed by different groups and on different substrates. The lack of knowledge regarding these apparently universal and basic phenomena demonstrates that the nucleation and formation of GaN NWs in PAMBE are still far from being completely understood.

Concerning the electronic and optical properties of \IDBs, it is helpful to imagine them as a thin type-II quantum well that binds holes. The Coulomb attraction exerted by these holes is strong enough to bind electrons, and the resulting \idb state decays radiatively and thus gives rise to intense light emission. The change in the symmetry of the fundamental hole state in \IDBs strongly modifies the polarization and the behavior of the \idb when subjected to magnetic fields. Donor atoms distributed in the vicinity of the \IDB plane localize the exciton within the \IDB plane. As a result of this localization, the \idb band resolves into a doublet with the lines at low and high energy being associated with localized and free \idb states, respectively. Their excitonic nature is manifested by the observation of a fine structure of the low energy line consisting of sharp, resolution-limited peaks. Finally, while the capture of excitons by \IDBs takes less than 50~ps, quasi-thermalization between the near-band edge excitons and the \idb takes much longer, resulting in the nonexponential decay usually observed for the \doxA transition in GaN NWs at low temperatures.

Analogously to stacking faults \cite{Corfdir2014b}, \IDBs do not suffer from fluctuations in layer thickness or composition as conventional quantum wells and thus offer unique possibilities for the study of low-dimensional excitons \cite{Corfdir2015}. Particularly interesting in this context are the tubular \IDBs formed in Ga-/N-polar core/shell NWs and intersections of these tubular \IDBs with stacking faults forming perfect crystal-phase quantum rings. Since the electronic states associated with all of these quantum structures are shallow \cite{Corfdir2014b}, they should be of little practical relevance for conventional devices. However, we envisage that their exceptionally well-defined properties make them ideal model systems for an understanding of quantum effects important for a future generation of optoelectronic devices.

\section{Acknowledgments}

The authors thank Pierre Lefebvre for fruitful discussions, Vincent Consonni and Caroline Chèze for providing additional samples, and Uwe Jahn for a critical reading of the manuscript. P. C. acknowledges funding from the Fonds National Suisse de la Recherche Scientifique through project 161032. This work was partly supported by the German BMBF joint research project MONALISA (Contract no. 01BL0810), by the Deutsche Forschungsgemeinschaft within SFB 951, and by Marie Curie RTN PARSEM (Grant No. MRTN-CT-2004-005583).


%

\end{document}